# CAPP Axion Search Experiments with Quantum Noise Limited Amplifiers

Sergey V. UCHAIKIN[1], Boris I. IVANOV[1], Jinmyeong KIM[2,1], Çağlar KUTLU[2,1], Arjan F. VAN LOO[3,4], Yasunobu NAKAMURA[3,4], Seonjeong OH[1], Violeta GKIKA[1], Andrei MATLASHOV[1], Woohyun CHUNG[1], Yannis K. SEMERTZIDIS[1]

[1]*Center for Axion and Precision Physics Research, IBS, Daejeon 34051, Republic of Korea*
[2]*Department of Physics, KAIST, Daejeon 34051, Republic of Korea*
[3]*RIKEN Center for Quantum Computing (RQC), Wako, Saitama 351–0198, Japan*
[4]*Department of Applied Physics, Graduate School of Engineering, The University of Tokyo, Bunkyo-ku, Tokyo 113-8656, Japan*

*\*E-mail: uchaikin@ibs.re.kr*

The axion is expected to solve the strong CP problem of quantum chromodynamics and is one of the leading candidates for dark matter. CAPP in South Korea has several axion search experiments based on cavity haloscopes in the frequency range of 1–6 GHz. The main effort focuses on operation of the experiments with the highest possible sensitivity. It requires maintenance of the haloscopes at the lowest physical temperature in the range of mK and usage of low noise components to amplify the weak axion signal. We report development and operation of low noise amplifiers for 5 haloscope experiments targeting at different frequency ranges. The amplifiers show noise temperatures approaching the quantum limit.

## 1. Introduction

Axions are one of the possible candidates for dark matter are axions. They are hypothetical elementary particles postulated by the Peccei-Quinn theory in 1977 to solve the strong CP problem in quantum chromodynamics [1,2]. Slow-moving axions produced in the early universe are also prospective dark matter candidates [3,4,5]. Axions interacting with a magnetic field can decay into two photons and thus be detected using an extremely sensitive receiver, the so-called axion haloscope. A haloscope consists of a high quality factor microwave cavity placed in a strong static magnetic field imposed by a superconducting magnet [6,7]. The signal power of the resulting microwave photons is very low and given by [8]

$$P_{a \to \gamma\gamma} \approx g_{a\gamma\gamma}^2 \frac{\rho_a}{m_a} B_0^2 V C \cdot \min(Q_L, Q_a) \sim 10^{-22} W, \qquad (1)$$

where $g_{a\gamma\gamma}^2$ is a model-dependent coupling constant, $\rho_a$ and $m_a$ are the axion density and mass, $B_0$ is the static magnetic field, $V$ is the effective haloscope volume, $C$ is the cavity form factor, and $Q_L, Q_a$ are the cavity and axion quality factors.

Because the axion mass is unknown, the resonance frequency of the cavity should be tunable to allow scanning all frequencies corresponding to all possible axion masses. The scanning speed is given by [9]

$$\frac{df}{dt} = \left(\frac{1}{SNR}\right)^2 \left(\frac{P_{a\to\gamma\gamma}(f)}{k_B T_{sys}}\right)^2 \frac{Q_a}{Q_L} \sim \frac{B_0^4 V^2 C Q_L}{T_{sys}^2}, \qquad (2)$$

where $SNR$ and $T_{SYS}$ are the read-out signal-to-noise ratio and system noise temperature, respectively. There have been numerous cavity experiments in search of axions (see, for example, [10] and [11]). To increase the scanning speed we need to increase the axion signal which is defined in the denominator of Eq. (2) and reduce the system noise temperature, which is in the numerator part of Eq. (2). A significant fraction of $T_{sys}$ is the noise temperature of the amplifier, $T_n$. At the early stage, only HEMT amplifiers were used at CAPP [12,13,14]. Theoretically, the lowest noise of any amplifier is given by the laws of quantum mechanics. Such an amplifier is said to have quantum-limited noise.

## 2. Flux Driven Josephson Parametric Amplifier

In CAPP we use flux-driven Josephson parametric amplifiers (FDJPA) with close to quantum-limited noise [15,16]. The FDJPA consists of a superconducting λ/4 resonator shorted to ground via a SQUID (Fig. 1). The central frequency of the resonator $f_c$ can be adjusted in a limited range by applying a dc flux using a superconducting coil. If a signal is applied on the FDJPA input, one can get amplification of the reflected signal. The energy for that process is provided by the pump tone applied to the SQUID. For our experiments, the parametric amplifiers are used in the non-degenerate three-wave mixing mode, and the frequency of the pump signal is close to twice the FDJPA resonance frequency.

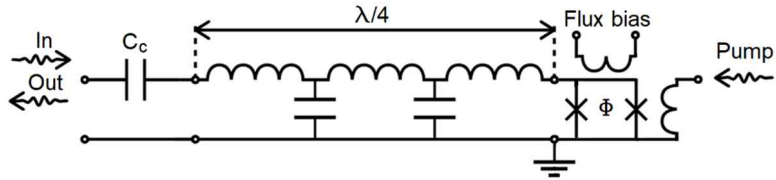

**Fig. 1.** Schematic of a flux-driven JPA.

## 3. Operation in the proximity of a strong magnet

Axion haloscope experiments use strong magnets to obtain the highest possible output signal. Devices based on Josephson junctions are very sensitive to external magnetic fields and should be protected from it. Each of the CAPP magnets has a compensation coil to provide a volume with reduced magnetic field, which is limited and highly occupied with RF components. To protect the FDJPA from the field of the 8T magnet, which is about 0.1 Tesla in the vicinity of the FDJPA, we developed a three-layer magnetic shield (Fig. 2). This shield was calculated to result in a residual magnetic field around an FDJPA of less than 100 nT. Figure 3 shows the dependence of the resonance frequency $f_c$ on the applied dc flux bias when the magnet is turned on and off. As one can see from Figure 3, the effect of the magnetic field on the resonance frequency is small.

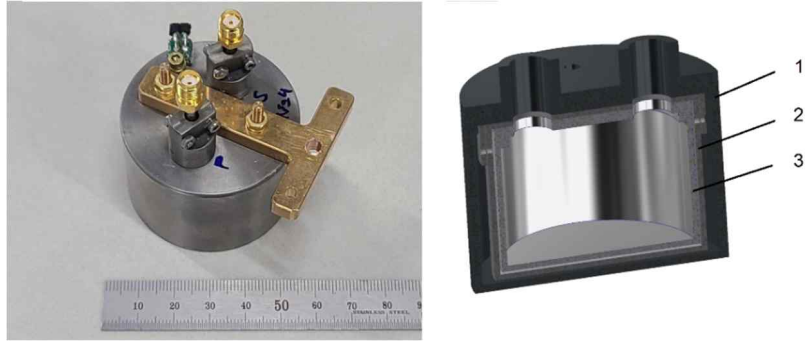

**Fig. 2.** Tree-layer FDJPA shield. 1 – NbTi layer; 2 – mu-metal layer; 3 – Al layer.

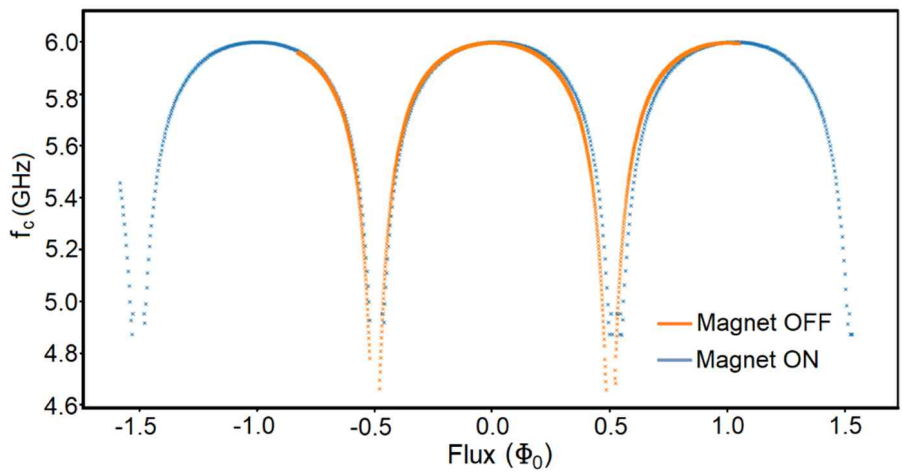

**Fig. 3.** Dependence of the resonance frequency $f_c$ on the applied dc flux bias when the 8 T magnet is turned ON and OFF. The residual field of the magnet at the position of the FDJPA without the shield is about 0.1 T. By implementing this shield, the field strength at the FDJPA is reduced to below 100nT.

## 4. FDJPA readout

A block diagram of the RF chain used for the axion experiment with a FDJPA as the first amplifier is shown in Fig. 4. The chain allows measuring different parameters such as tuning frequency range, gain, and instantaneous bandwidth of the FDJPA. The PUMP input provides the pump signal of the FDJPA. The output signal of the FDJPA is further amplified with low-noise cold HEMT amplifiers. The setup also has a noise source which allows for measuring of noise temperature of the FDJPA and HEMT. The cold RF switch is used to connect the FDJPA to the noise source or the cavity. To measure the cavity and the RF chain properties the WEAK, BYPASS and CAVR inputs are used. The WEAK input is connected with a special cavity antenna weakly coupled to the cavity volume. The BYPASS input allows us to measure the RF chain characteristics by bypassing the cavity. CAVR is used to obtain the cavity reflection (for more details, see [16,17]).

## 5. Implementation of FDJPAs into axion experiments

In the CAPP flagship "12TB" and planned "18T" experiments we use wet dilution fridges which consume a lot of liquid helium during the cooldown of our equipment. To save budget and minimize the number of cooldowns we made a special design of the mixing chamber RF assembly (Fig. 5).The assembly combines all of the cold RF components: FDJPA, circulators, couplers, switch, noise source, etc. The assembly can be mounted in BlueFors dry fridges for preliminary testing without using liquid helium.

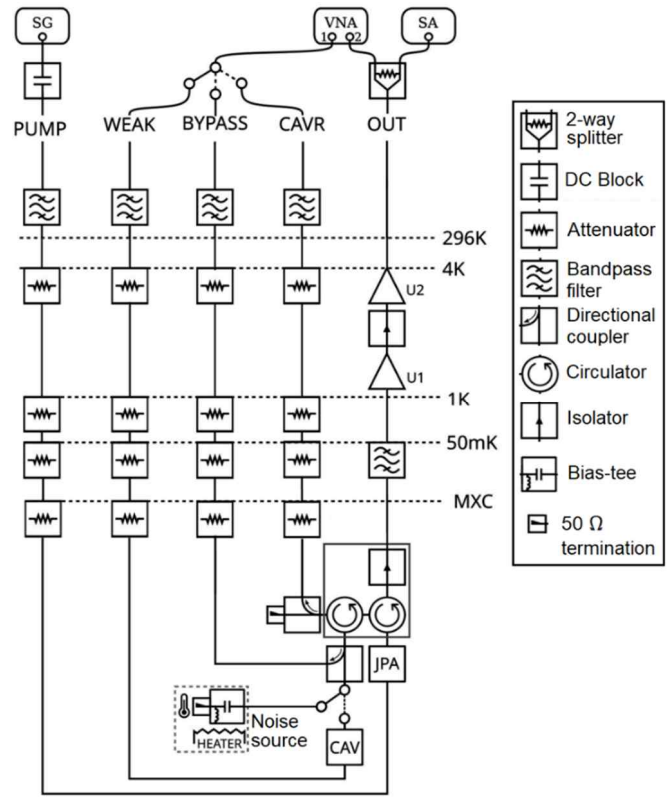

**Fig. 4.** Block diagram of the microwave setup in axion experiments. SG, VNA and SA denote a signal generator, vector network analyzer and spectrum analyzer, respectively. PUMP, WEAK, BYPASS, CAVR and OUT are RF connectors on the top of the fridge. U1 and U2 are low temperature HEMT amplifiers. MXC is the mixing-chamber plate, and CAV is the cavity.

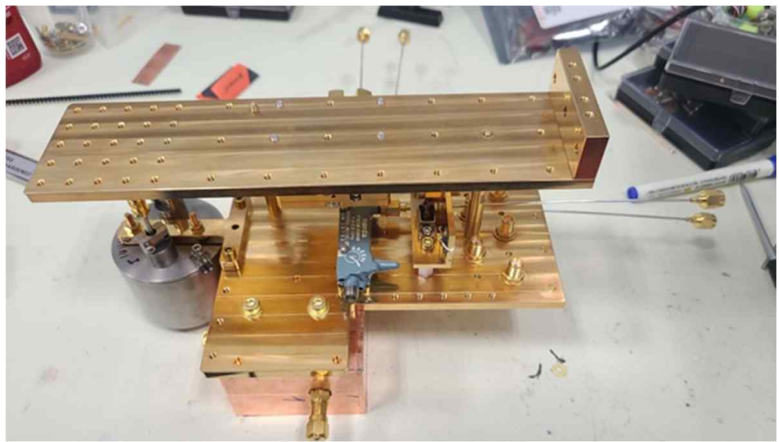

**Fig. 5.** MXC assembly of the 0.95–1.35 GHz cold RF chain for the CAPP 12TB experiment.

**Table I.** Selected FDJPAs for CAPP axion search experiments. $T_n$, $hf/(2k_B)$ and $T_n/(hf/(2k_B))$ are the noise temperature of the FDJPA, the quantum-limited noise for amplifier added noise and the ratio between them for the middle of the FDJPA operating frequency range.

| JPA No. | 2010-1.05G | 2010-1.15G | 2010-1.1G | 2010-1.2G | 2010-1.25G | 2010-2.25G | 1904-2.4G | 1910-2.5G | 1910-6.09G | 1910-7G |
|---|---|---|---|---|---|---|---|---|---|---|
| Operation Frequency Low-High (GHz) | 0.98–1.018 | 1.006–1.071 | 1.07–1.12 | 1.13–1.165 | 1.17–1.21 | 2.06–2.157 | 2.27–2.31 | 2.3–2.45 | 5.1–5.4 | 5.5–6.0 |
| Tunable range (MHz) | 20 | 65 | 50 | 35 | 40 | 97 | 40 | 15 | 300 | 500 |
| $T_n$ (mK) | 100 | 90 | 100 | 95 | 100 | 150 | 120 | 130 | 145 | 150 |
| $hf/(2k_B)$ | 24 | 25 | 26 | 28 | 29 | 51 | 55 | 57 | 126 | 138 |
| $T_n/(hf/(2k_B))$ | 4.2 | 3.6 | 3.8 | 3.4 | 3.4 | 3 | 2.2 | 2.28 | 1.15 | 1.09 |
| Bandwidth (kHz) | >150 | >150 | >150 | >150 | >200 | >200 | ~100 | ~100 | ~1000 | ~1000 |

We have 25 FDJPAs for a frequency range 0.98–6.0 GHz. Some of them are shown in Table 1 and Figure 6. As you can see from Figure 6, the noise temperature varies for different amplifiers and operating frequency. We assume that this is the influence of the RF circuits and are working to improve them. Figure 7 shows the coverage of the frequency for a running 12TB 1.1 GHz with available FDJPAs. The FDJPAs have been used in completed [17] and ongoing CAPP experiments: 8T PACE 2.3 GHz, 8T PACE 5.6 GHz, 8TB 5.9 GHz, 12TB 1.1 GHz and 8T PACE 2.3 GHz with a superconducting cavity. Results of these experiments will be reported soon.

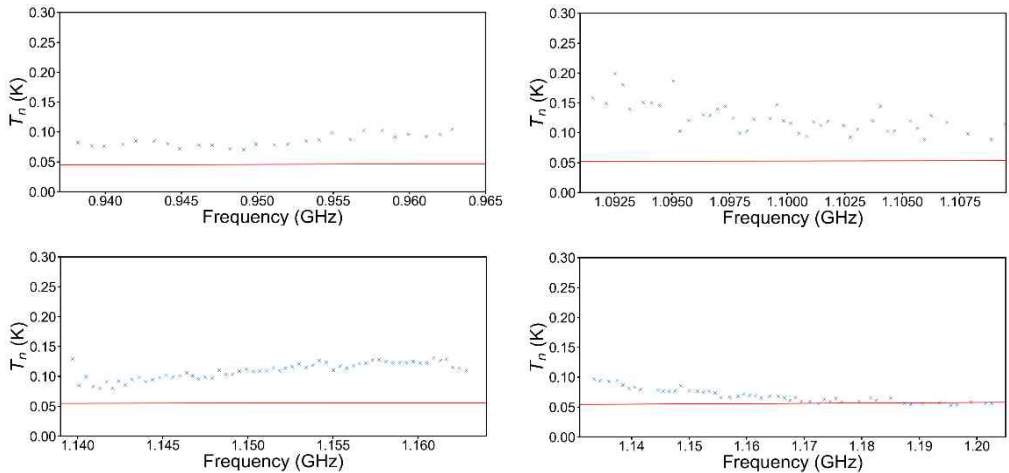

**Fig. 6.** Results of noise measurements of 4 FDJPAs. The red line corresponds to the standard quantum limit $hf/k_B$.

## 6. Conclusion

We developed a set of FDJPAs for a frequency range of 0.98 to 6 GHz as well as methods and designs to implement them in CAPP axion search experiments. Noise temperatures of all amplifiers approach the quantum noise limit and their gain (>20 dB) allows minimizing the influence of subsequent amplifying stages on the system noise temperature. The FDJPAs have so far been used in 5 completed and ongoing CAPP experiments.

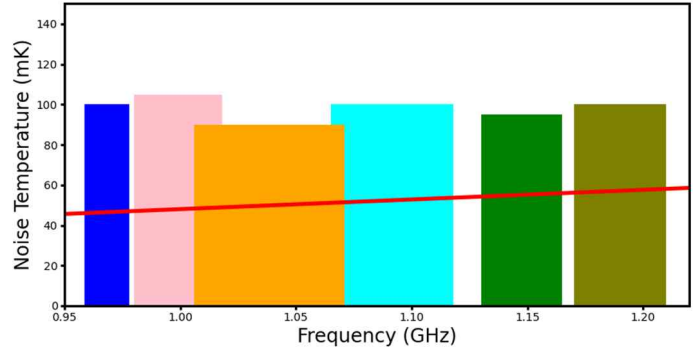

**Fig. 7.** Coverage of 0.95–1.2 GHz MXC assembly of 7 FDJPAs. Each bin correspond to one FDJPA. The bin width and height show the FDJPA tunable bandwidth and noise temperature. The red line corresponds to the standard quantum limit $hf/k_B$.

## Acknowledgment

This work is supported by IBS-R017-D1-2020-a00 and JST ERATO (Grant No. JPMJER1601). Arjan F. van Loo was supported by the JSPS postdoctoral fellowship.